\documentclass{elsart}

\usepackage{graphicx}

\usepackage{amssymb}

\usepackage{float}

\newcommand{\flr}[3]{\mathopen{}\lr#1{#2}#3\mathclose{}}
\newcommand{\lr}[3]{\left#1#2\right#3}
\renewcommand{\d}[0]{{\rm d}}

\begin{document}

\begin{frontmatter}



\title{Photon emission by ultra-relativistic positrons
in crystalline undulators: the high-energy regime.\thanksref{contrib}}

\thanks[contrib]{contributed to the 23rd International Free Electron Laser
Conference, Darmstadt, Germany, 20-24 August 2001
}

 
\author{W. Krause\thanksref{email}}
\author{A. V. Korol}
\author{A. V. Solov'yov}
\author{W. Greiner}

\address{Institut f\"ur Theoretische Physik der Johann Wolfgang
Goethe-Universit\"at, Postfach 11 19 32, 60054 Frankfurt am Main, Germany}

\thanks[email]{E-mail: krause@th.physik.uni-frankfurt.de}

\begin{abstract}
This paper discusses the undulator radiation emitted by high-energy
positrons during planar channeling in periodically bent crystals.  We
demonstrate that the construction of the undulator for positrons with
energies of 10 GeV and above is only possible if one takes into
account the radiative energy losses.  The frequency of the undulator
radiation depends on the energy of the particle. Thus the decrease of
the particle's energy during the passage of the crystal should result
in the destruction of the undulator radiation regime.  However, we
demonstrate that it is possible to avoid the destructive influence of
the radiative losses on the frequency of the undulator radiation by
the appropriate variation of the shape of the crystal channels. We
also discuss a method by which, to our mind, it would be possible to
prepare the crystal with the desired properties of its channels.
\end{abstract}

\begin{keyword}
undulator radiation \sep channeling positrons \sep
periodically bent crystal \sep tapered undulator \sep
strained crystals
\PACS 41.60.-m
\end{keyword}
\end{frontmatter}


\section{Introduction}
\label{sec:intro}

In this paper we discuss a mechanism, initially proposed in
\cite{Korol98,Korol99}, for the generation of high-energy photons
by means of planar channeling of ultra-relativistic positrons through
a periodically bent crystal. In addition to the well-known channeling
radiation, there appears an undulator type radiation due to the
periodic motion of the channeling positrons which follow the bending
of the crystallographic planes. The intensity and the characteristic
frequencies of this undulator radiation can be easily varied by
changing the positrons energy and the parameters of the crystal
bending.

\begin{figure}
\includegraphics{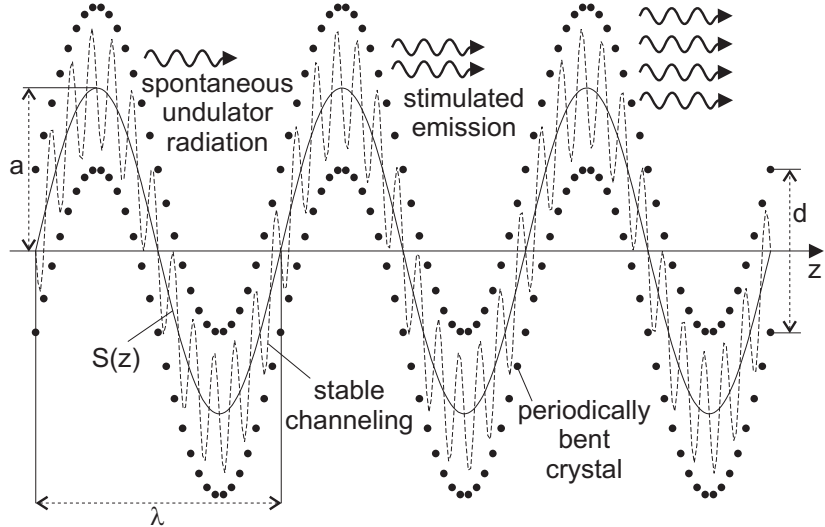}
\caption{Schematic figure of the crystalline undulator. The scale in
$y$ direction is magnified by a factor $> 10^4$. The function $S(z)$
defines the shape of the centerline of the periodically bent channel.}
\label{fig:undulator}
\end{figure}

The mechanism of the photon emission by means of the crystalline
undulator is illustrated in figure \ref{fig:undulator}.  It is
important to stress that we consider the case when the amplitude
$a$ of the bending is much larger than the interplanar spacing $d$
($\sim 10^{-8}\;\mathrm{cm}$) of the crystal ($a \sim 10\ d$), and,
simultaneously, is much less than the period $\lambda$ of the
bending ($a \sim 10^{-5} \dots 10^{-4}\, \lambda$).

In addition to the spontaneous photon emission the scheme leads to the
possibility to generate stimulated emission. This is due to the fact,
that the photons emitted at the points of maximum curvature of the
trajectory travel almost parallel to the beam and thus, stimulate the
photon generation in the vicinity of all successive maxima and minima
of the trajectory.

We consider two methods for bending the crystal. First, a
transverse acoustic wave could be used to bend the crystal dynamically
\cite{Korol98,Korol99}. One possibility to create acoustic waves in a
crystal is to place a piezo sample atop the crystal and to use radio
frequency to excite oscillations. Second, a statically and
periodically bent crystal could be constructed using graded strained
layers \cite{Uggerhoj00}. We will present a detailed description how a
static crystalline undulator can be produced.

The conditions for stable channeling has been discussed in
\cite{Korol99,Korol00,Krause00a}. Channeling only takes place if the maximum
centrifugal force in the bent channel, $F_{\mathrm{cf}}\approx m \gamma
c^2/R_{\mathrm{min}}$ ($R_{\mathrm{min}}$ being the minimum curvature
radius of the bent channel), is less than the maximal force due to the
interplanar field, $F_{\mathrm{int}}$ which is equal to the maximum
gradient of the interplanar field (see \cite{Korol99}). As shown in
\cite{Korol00} the ratio $C=F_{\mathrm{cf}}/F_{\mathrm{int}}$ has to be
smaller than 0.15, otherwise the phase volume of channeling
trajectories is too small. Thus, the inequality $C<0.15$ connects the
energy of the particle, $\varepsilon=m \gamma c^2$, the parameters of
the bending (these enter through the quantity $R_{\mathrm{min}}$), and
the characteristics of the crystallographic plane.

Another important effect is the dechanneling of the projectiles.  The
channeling particles scatter with electrons and nuclei of the
crystal. These random collisions lead to a gradual increase of the
particle energy associated with the transverse oscillations in the
channel. As a result, the transverse energy at some distance $L_d$
from the entrance point exceeds the depth of the interplanar potential
well, and the particle leaves the channel. The quantity $L_d$ is
called the dechanneling length \cite{Gemmel74} and can be calculated
using the methods described in \cite{Krause00a,Biryukov96}. Thus,
to consider the undulator radiation formed in a crystalline undulator,
it is meaningful to assume that the crystal length does not exceed
$L_d$.

In \cite{Krause00a} we estimated the parameters $a$ and $\lambda$ for
given energy $\varepsilon$, regarding the dechanneling length of the
bent crystal and the reduction of the phase-space volume due to the
bending.  For 500 MeV positrons in Si(110) the optimal parameters are
$a/d=10$ and $\lambda = 2.335 \cdot 10^{-3}$ cm. The spectral
distribution of the emitted radiation in this case is discussed in
the next section (see also \cite{Krause00a}).

In the present paper we discuss the possibility to construct
undulators to generate photons with energies larger than 1 MeV using
positron energies above 10 GeV when the radiative energy losses cannot
be neglected and, thus, must be taken into account \cite{Korol00}.

The frequency of photons generated in the undulator is determined by
the energy of the projectiles and also by the undulator parameter (for
definition see equation (\ref{losses_comp_1})). In the regime in which
the energy of the projectiles is not constant during their passage
through the undulator, the frequency of the emitted undulator
radiation can nevertheless be kept constant if one chooses the
appropriate variation of the shape of the undulator along its length.

We also discuss a method by which, to our mind, it would be possible
to prepare crystals with the desired properties of their channels.

\section{Spectra of the spontaneous emitted radiation}
\label{sec:spectra}

To illustrate the undulator radiation phenomenon, which we discuss,
let us consider the spectra of spontaneous radiation emitted during
the passage of positrons through periodically bent crystals.

The photon emission spectra have been calculated using the
quasiclassical method \cite{Baier98}.  The trajectories of the
particles were calculated numerically and then the spectra were
evaluated \cite{Krause00a}. The latter include both radiation
mechanisms, the undulator and the channeling radiation.

\begin{figure}
\includegraphics{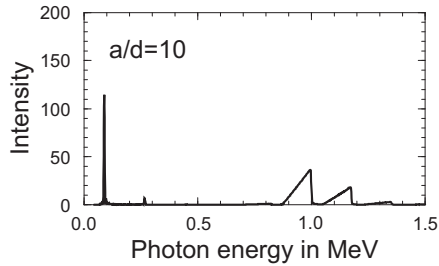}
\caption{Spectral distributions of the total radiation emitted in
forward direction for $\varepsilon=500$ MeV positrons channeling in Si
along the (110) crystallographic planes for $a/d=10$.}
\label{fig:spectrum_500mev}
\end{figure}

Figure \ref{fig:spectrum_500mev} shows the spectral distribution of
the total radiation emitted in forward direction for $\varepsilon=500$
MeV positrons channeling in Si along the (110) crystallographic
planes. The wavelength of the periodically bent crystal is
$\lambda=2.335\cdot10^{-3}$ cm and its amplitude is $a=10\,d$. For the
length of the crystal we chose $L_d=3.5\cdot10^{-2}$ cm. This
corresponds to $N=15$ undulator periods.

The first and third harmonic of the undulator radiation are visible on
the left side of the spectrum (about 100 keV). The large triangular
peak around 1 MeV corresponds to the first harmonic of the channeling
radiation. The smaller triangular peaks appear due to the coupling of
undulator and channeling radiation. The two types of
radiation are well separated and the influence from the channeling
radiation on the undulator radiation is negligible small. A detailed
discussion of this spectrum is given in \cite{Krause00a}.

\section{Undulator effect in the high-energy regime}
\label{cha:losses}

The spectra of the emitted radiation presented in the
previous section has been calculated in the regime in which the energy
losses of the positrons during their passage through the crystal are
negligible. In this section, we analyze the opposite situation, which
occurs when the energy of the projectiles becomes larger than 10 GeV. On the first glance, the undulator phenomenon can
hardly take place in this energy range, because the energy of positrons
during their passage through the crystal can no longer be considered
as constant due to the radiative energy losses \cite{Korol00}.

Indeed, the frequency $\omega_{\mathrm{und}}^{(1)}$ of the first
harmonic of the undulator radiation in the forward direction is given
by \cite{Korol99,Korol01a}:
\begin{equation}
\label{losses_comp_1}
\omega_{\mathrm{und}}^{(1)}=
\frac{4\,\omega_0\,\gamma^2}{2+p_{\mathrm{und}}^2}
= \frac{4 \pi\, c \, \gamma^2(z)}{\lambda+2\pi^2 \,\frac{a^2}{\lambda}\,
\gamma^2(z)}.
\end{equation}
Here we use $\omega_0=2\pi\,c/\lambda$ and the undulator parameter
$p_{\mathrm{und}}$ is defined as $p_{\mathrm{und}}= \gamma \, 2 \pi\,
a/\lambda$. The shape of the crystal is $S(z)=a\, \sin(kz)$.

Equation (\ref{losses_comp_1}) shows that the frequency of the emitted
radiation depends on the energy of the projectile. If the decrease of
the particle's energy due to the radiative losses is significant
($\gamma(z) < \gamma_0$ for $z>0$), the frequency
$\omega_{\mathrm{und}}^{(1)}$ becomes dependent on the particle's
penetration distance $z$ into the crystal. The decrease of the
particle's energy leads to the broadening of the undulator lines in
the photon emission spectrum and the reduction of their intensity.

However, the monochromaticity of the undulator radiation in the
high-energy regime can be restored if one allows the variation of the
shape of the crystal channels. Let us consider this condition in more
detail and assume that the shape of the channels in the crystal is as
follows:
\begin{equation}
\label{losses_comp_3}
S(z)=a(z) \, \sin\flr({\varphi(z)})
\end{equation}
with $\varphi(z)=\int_0^z\, 2\pi/\lambda \, \d z$,  $a(z)$ and
$\lambda(z)$ are the amplitude and the ``period'' of the bent
crystal channels as function of the penetration depth
$z$.

Let us formulate the conditions for the choice of the shape function
$S(z)$. For the given dependence $\gamma(z)$ the functions $a(z)$ and
$\lambda(z)$ have to be chosen to keep constant the frequency of the
first harmonics, $\omega_{\mathrm{und}}^{(1)}(z) =
\mathrm{const.}$ In addition, we require $C(z)=C=\mathrm{const.}$ It
was shown in \cite{Krause00a,Korol01a}, that the parameter
$C=\varepsilon(z)/(R_{\mathrm{min}}(z)\cdot
U_{\mathrm{max}}^\prime)$ is characteristic for the
channeling process in bent channels and the regime in which the
process happens. Here, $R_{\mathrm{min}}\approx
\lambda^2(z)/(4\pi\,a(z))$ is the curvature radius of the shape
function $S(z)$ in the points of its extrema.

The two conditions $\omega_{\mathrm{und}}^{(1)}(z) = \mathrm{const.}$
and $C(z)=C=\mathrm{const.}$ and the knowledge of $\gamma(z)$ allow
the calculation of $a(z)$ and $\lambda(z)$. Thus the shape $S(z)$ of
the crystal is known. For comparatively low energies of the projectile
($\varepsilon < 10$ GeV) the dependence $\gamma(z)$ can be calculated
using the approach suggested in \cite{Korol00}. To describe the
radiative losses of particles in the high-energy regime, one has to
modify the formulas outlined in \cite{Korol00} and consider the
continuously decreasing energy.

Thus, using the initial values $\varepsilon_0$, $a_0$ and $\lambda_0$,
one can calculate the energy $\varepsilon(z)$ as function of the
penetration distance $z$.  Using this dependency we then can calculate
the shape $S(z)$ of the channel.  The latter, in turn, ensures that
the frequency of the undulator radiation and the parameter $C$ remain
constant during the passage of the positrons through the crystal, even
in the regime in which the radiative energy losses are high. We
consider the possibility of the construction of such bent crystals in
the next section.

\begin{figure}
\includegraphics{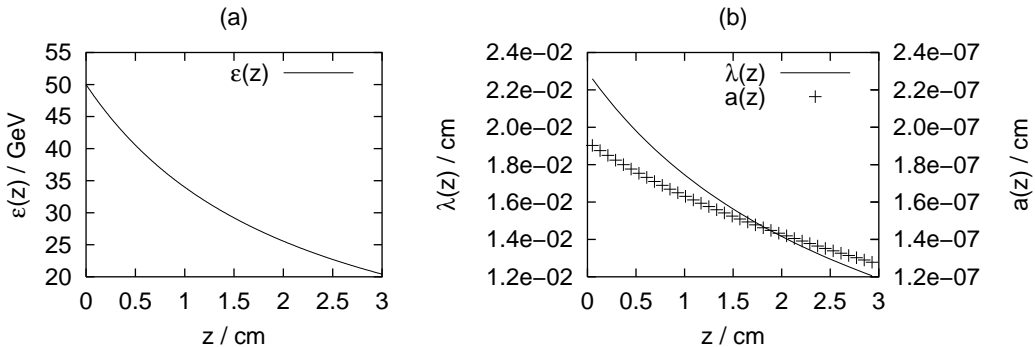}
\caption{(a) The energy of the positrons as function of the penetration
depth $z$ in the high-energy channeling regime for Si(110) and initial
positron energy 50 GeV. The averaging over the possible initial
conditions of the positrons was performed as described in
\cite{Korol00}.
(b) The wavelength $\lambda$ and the amplitude $a$ of the bent
crystal as function of the penetration depth $z$.
}
\label{fig:losses}
\end{figure}

To illustrate the described method we consider positrons with an
initial energy of 50 GeV channeling in Si(110). Figure
\ref{fig:losses} (a) presents the energy of positrons as a
function of the penetration depth. We have chosen $C=0.15$ and the
initial amplitude $a_0= 10\, d$. These parameters define $\lambda_0 =
\sqrt{\varepsilon \, \pi^2\, a\, d/(U_{\mathrm{max}}^\prime\, C)} =
2.25\cdot 10^{-2}$ cm. The argumentation for the choice of $C$ and $a$
one finds in \cite{Korol00} and \cite{Krause00a}.  The calculated
values for $\lambda(z)$ and $a(z)$ are presented in figure
\ref{fig:losses} (b). Knowing these dependencies, one can easily
calculate the shape of the channels.

The particle density of channeling positron beams decreases exponentially
along the channel \cite{Biryukov96,Korol01a}.  The dechanneling length
for positrons of $\varepsilon=50$ GeV and $C=0.15$ is approximately
1.46 cm and the number of undulator periods on this length is about
75. The emitted undulator radiation should have
high intensity and narrow spectral width. The energy of photons at
the first harmonic emitted in the forward direction is
$\hbar\omega_{\mathrm{und}}^{(1)}=7.1$ MeV and the spectral width can
be estimated as $\Gamma_{\mathrm{und}}^{(1)}/2 =
\hbar\omega_{\mathrm{und}}^{(1)} / N_{\mathrm{und}} = 44$ keV.

\section{Growing of crystals with periodically bent channels}
\label{cha:crystal}

In this section we propose a method of preparing crystals with
periodically bent channels whose shape function $S(z)$ has either the
pure sine form, $a\, \sin kz$, or a more general one defined by
(\ref{losses_comp_3}). In \cite{Breese97c} the deflection of proton
beams by means of strained crystal layers was demonstrated. The
construction of the crystals was described and experimental data that
proves the deflection of protons was presented.

\begin{figure}
\includegraphics{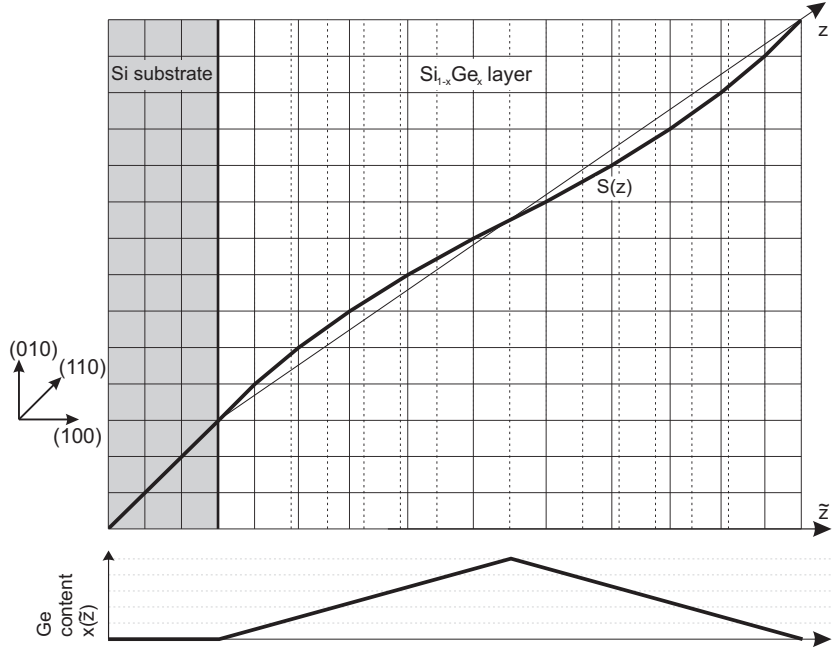}
\caption{Schematic figure of the first period of the bent crystal.}
\label{fig:bent_crystal_first_period}
\end{figure}

Using well-known methods of crystal growing (like molecular beam
epitaxy or chemical vapour deposition, see the references in
\cite{Breese97c}) it is possible to add single crystal layers onto a
substrate. Let us consider a pure silicon substrate on which a
$\mathrm{Si}_{1-x}\mathrm{Ge}_x$ layer is added ($x$ denotes the
germanium content in this layer). The doping with germanium leads to
the enlargement of the lattice constant of the added layer. The strain
due to the lattice mismatch of the substrate and the
$\mathrm{Si}_{1-x}\mathrm{Ge}_x$ layer leads to an increase of the
lattice spacing perpendicular to the surface of the substrate (the
$\tilde z$-direction in figure
\ref{fig:bent_crystal_first_period}). The lattice constant parallel to
the surface remains unchanged.

Prior to discussing the growing of periodically bent channels, let us
summarize the main ideas presented in \cite{Breese97c} that we need
for our description. The spacing between the (100) layers is
$d_{\mathrm{Si}}=1.358$ \AA\ in Si and $d_{\mathrm{Ge}}=1.414$ \AA\ in
Ge. The distance between two $\mathrm{Si}_{1-x}\mathrm{Ge}_x$ layers
is given by $d(x)=d_{\mathrm{Si}}+ \Delta d \cdot x$, where $\Delta
d=d_{\mathrm{Ge}} - d_{\mathrm{Si}}$. In \cite{Breese97c} the critical
thickness of the strained layer is discussed. If the thickness of the
strained layer is larger than the critical value $h_c$, then lattice
defects appear and destruct the channels.

To obtain periodically bent channels, one starts with a pure silicon
substrate and adds $\mathrm{Si}_{1-x}\mathrm{Ge}_x$ layers with
continuously increasing Ge content. This results in bending of the
(110) channels in the direction of the (100) channels. The periodicity
of the shape requires the change of the direction of the bending
toward the (010) channels. This, in turn, can be achieved by reducing
$x$ until it reaches 0. Figure \ref{fig:bent_crystal_first_period}
schematically illustrates the first period of the bent (110) channel.

The last crystal layer of the first period consists of pure
silicon, so that the second period can be built up on top of the first
in the same manner. To be captured by the bent channel, the positron
beam should be directed towards the (110) channel of the substrate.

The crystal strain is strongest after half a period, when the
germanium content reaches its maximum. The thickness of the layers
corresponding to half a period needs to be smaller than the critical
thickness $h_c$. If this condition is met, then crystals with
arbitrary number of undulator periods can be constructed.

For a given shape $S(z)$ it is possible to calculate the germanium
content $x(\tilde z)$ as a function of the thickness $\tilde z$ (see
figure \ref{fig:bent_crystal_first_period}). The formulas can be
derived using simple geometric considerations and are discussed in
\cite{Krause01a}. We now consider two examples.

\begin{figure}
\includegraphics{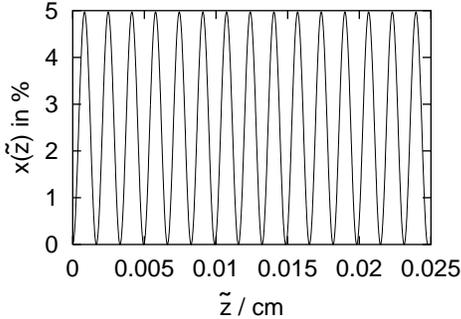}
\caption{Germanium content as function of $\tilde z$ for a bent
Si(110) crystal. The parameters are equal to those used for the
calculation of the spectrum shown in figure \ref{fig:spectrum_500mev}.}
\label{fig:crystal_500mev_x}
\end{figure}

First we discuss growing the Si crystal with the sine-like shape
$S(z)= a\, \sin kz$ with $a=10\,d=1,92\cdot 10^{-7}$ cm and
$\lambda=2\pi/k = 2.335\cdot 10^{-3}$ cm.  These parameters correspond
to the undulator emission spectrum presented in figure
\ref{fig:spectrum_500mev} for $a/d=10$. The germanium content obtained
by our calculations is plotted in figure
\ref{fig:crystal_500mev_x}.

The maximum germanium content is 5\%. The layer thickness that
corresponds to half a period is given by $\lambda/(2\,\sqrt{2})
=0.8\cdot 10^{-3}$ cm. The critical thickness $h_c$ for a strained
crystal with 5\% of Ge is about $1.2\cdot 10^{-3}$ cm
\cite{Breese97c}.

The second example concerns the shape function given by
(\ref{losses_comp_3}) with $a(z)$ and $\lambda(z)$ as in figure
\ref{fig:losses} (b). To find the dependence $x(\tilde
z)$ in this case is not so straightforward as for the sine
profile. The calculation of $x(\tilde z)$ results in negative
values. To avoid this problem one can consider the crystal growth in
the inverse direction: $S(z) \rightarrow S(L_d -z)$ for $0 \le z \le
L_d$. Then the calculated germanium content is positive for all
$\tilde z>0$. The projectiles are injected not through the substrate,
as in the first example, but from the opposite side of the
crystal. For more details and the analytic description see
\cite{Krause01a}.

\section{Summary and outlook}

In this work we have discussed the high-energy regime of the undulator
radiation emitted by ultra-relativistic positrons channeling in
periodically bent crystal channels.

This regime is typical for positron energies well above 10 GeV, when
the channeling effect is accompanied by noticeable radiative
losses. The latter, being mainly due to the channeling radiation, lead
to the gradual decrease of the positron energy. This, in turn,
strongly influences the stability of the parameters of the emission of
undulator radiation.

We demonstrated that the frequency of the undulator radiation can be
maintained constant provided the parameters of the periodic bending
are changed with the penetration distance to take into account the
decrease of the projectile energy.

Our investigation shows that the discussed modification of the shape
of the crystal channels allows the generation of undulator radiation
of high-energy photons (up to tens of MeV). The calculation of the
spectral distributions of the emitted photons in this regime is
currently in progress and will be reported soon.

We described a method that should allow the growing of the crystal
channels that are necessary for the experimental measurement of the
photon spectra. The feedback from experimentalists would be very
helpful to check the predictions, models and assumptions that were
used in this work.

\ack
The research was supported by DFG, BMBF and the Alexander von Humboldt
Foundation.

\bibliography{fel2001-Tu-O-14}

\end{document}